\documentclass[aip,reprint,prb,amsmath,amssymb]{revtex4-1}

\usepackage{graphicx}% Include figure files
\usepackage{dcolumn}% Align table columns on decimal point
\usepackage{bm}% bold math
\usepackage{subfig}
\usepackage[utf8]{inputenc}
\usepackage{amsmath,amssymb}
\usepackage{epstopdf}
\usepackage{textcomp}
\usepackage{hyperref}
\usepackage{listings}
\usepackage{upgreek}

\newcommand{\be}{\begin{equation}}
\newcommand{\ee}{\end{equation}}
\newcommand{\beal}{\begin{align}}
\newcommand{\enal}{\end{align}}
\newcommand{\bear}{\begin{eqnarray}}
\newcommand{\eear}{\end{eqnarray}}
\newcommand{\nn}{\nonumber}

\newcommand{\e}{\mathrm{e}}

\begin{document}

\title{Slow magnetosonic wave absorption
by pressure induced ionization-recombination dissipation}

\author{Todor~M.~Mishonov}
\email[E-mail: ]{mishonov@bgphysics.eu}
\affiliation{Institute of Solid State Physics, Bulgarian Academy of Sciences,
72 Tzarigradsko Chaussee Blvd., BG-1784 Sofia, Bulgaria}

\author{Albert~M.~Varonov}
\email[E-mail: ]{varonov@issp.bas.bg}
\affiliation{Institute of Solid State Physics, Bulgarian Academy of Sciences,
72 Tzarigradsko Chaussee Blvd., BG-1784 Sofia, Bulgaria}

\date{15 May 2020, 21:05}

\begin{abstract}
A new mechanisms for damping of slow magnetosonic waves (SMW) by pressure induced oscillations of the ionization degree is proposed. 
An explicit formula for the damping rate is quantitatively derived. 
Physical conditions where the new mechanism will dominate are briefly discussed.
The ionization-recombination damping is frequency independent and has no hydrodynamic interpretation.
Roughly speaking large area of partially ionized plasma
are damper for basses of SMW while usual MHD mechanisms operate as a low pass filter.
The derived damping rate is proportional to the square
of the sine between the constant magnetic field 
and the wave-vector.
Angular distribution of the spectral density of
SMW and Alfv\'en waves (AW)
created by turbulent regions and passing 
through large regions of partially ionized plasma
is qualitatively considered. 
The calculated damping rate is expressed by the
electron impact cross section of the Hydrogen atom
and in short all details of the proposed damping mechanisms are well studied.
\end{abstract}

\maketitle

%%%%%%%%%%%
\section{Short Introduction}

Behind purely fundamental interest for plasma physics 
propagation of hydromagnetic (nowadays known as magnetohydrodynamic ) waves attracted significant attention 
and was strongly simulated by the development of the physics of solar atmosphere
and the eternal problems related to its heating.\cite{Alfven:42,Alfven:47,LL8}
It has already been confirmed that the magnetohydrodynamic (MHD) waves (both incompressible and compressible) are present in the solar atmosphere and they have already been considered for heating of the solar chromosphere and corona.\cite{Mond:94,Campos:99,Wijn:07,Ruderman:12}
Models adapted to study heating problems of solar atmosphere include
two fluids coupled through collisions and chemical reactions, such as
impact ionization and radiative recombination with imposed initial thermal and chemical equilibrium.
Within this approach the plasma heating is dominantly wave-based and the main energy source for heating are the excited fast magnetosonic fluctuations,\cite{Maneva:17} while older studies of FMW heating can be found in Ref.~\onlinecite{Zhelyazkov:87} for instance. 

It is worthwhile to mention also works on overreflection or swing amplification
in shear flow of slow magnetosonic waves (SMW); see for example Refs.~\onlinecite{Hristov:11,Gogoberidze:04}.

In a review on partially ionized astrophysical plasmas\cite{Ballester:18}
it is shown that viscosity plays no important role in the damping of chromospheric Alfv\'en waves and recently it is concluded that the solar corona electrical resistivity has only very small impact, while and thermal conduction and viscosity contribute equally.\cite{Perelomova:20}
Therefore, the question of chromospheric heating due to the ion-neutral interaction will require further studies in the future.
A complete review on the problem requires many hundred citations,
but here we mention the importance  of two fluid approach for consideration of
MHD waves in partially ionized plasmas.\cite{Zaqarashvili:11} 

%%%%%%%%%%%%%%%

\subsection{Scenario}

When a slow magnetosonic wave (SMW) propagates
through partially ionized plasma,
the oscillations of the pressure creates oscillations of the temperature
and generates small oscillations of the degree of ionization 
$\alpha$. 
Those pressure induced deviations of the chemical equilibrium gives an extra entropy production and energy dissipation of the SMW.

This additional mechanism does not work for the 
Alfv\'en waves (AW) and in spite of common dispersion
and damping of AW and SMW in (MHD) approach at small magnetics field their 
ionization-recombination absorption can be completely different.

The purpose of the present work is to present an explicit formula for the chemical damping and to consider in short when the predicted new damping mechanism 
is important and dominates  and how it can be observed.

The article is organized as follows. In order to create the necessary system of notions
and notations following Landau and Lifshitz\cite{LL8}  
in the next Sec.~\ref{Recalling SMW} 
we will recall the physics of SMW.
Then we derive in 
Sec.~\ref{Ionization-recombination absorption} 
our new result for ionization-recombination absorption.
Finally we will discus in 
Sec.~\ref{Discussion and conclusions}

%%%%%%%%%%%%%
\section{Recalling SMW}
   \label{Recalling SMW}
In this section we will repeat these details which are common for Alfv\'en waves (AW) and SMW.
The differences between AW and SMW which are 
our new result we derive after that.

%%%%%%%%%%%
\subsection{Dispersion of MHD Waves}
Low density hydrogen plasma we approximate 
as a cocktail of ideal gases of electrons, protons and neutral atoms with pressure $p$ and mass density
$\rho$
\beal
& p=nT, \qquad n=n_e+n_p+n_0,\\
& \rho=n_\rho M, \qquad n_\rho=n_p+n_0,
\end{align}
where temperature $T$ is written in energy units 
and $M$ is the proton mass. 
The sound velocity is defined by the adiabatic compressibility 
\be
c_s=\sqrt{\left(\frac{\partial p}
{\partial \rho}\right)_{\!\!s}},
\ee
for which the standard expression
from the averaged atomic mass of the cocktail
$\left<M\right>$
\beal
&
c_s=\sqrt{\frac{\gamma_p T}{\left<M\right>}},\quad
\left<M\right>=\frac{n_p M+n_0 M+n_e m}
{n_p+n_0+n_e},\\
&
c_v=3/2, \quad c_p=c_v+1= 5/2,\quad
\gamma=c_p/c_v=5/3
\end{align}
where $c_v$ and $c_p$ are the heat capacities 
per atom and $m$ is the electron mass.

\begin{figure}[h]
\includegraphics[scale=0.6]{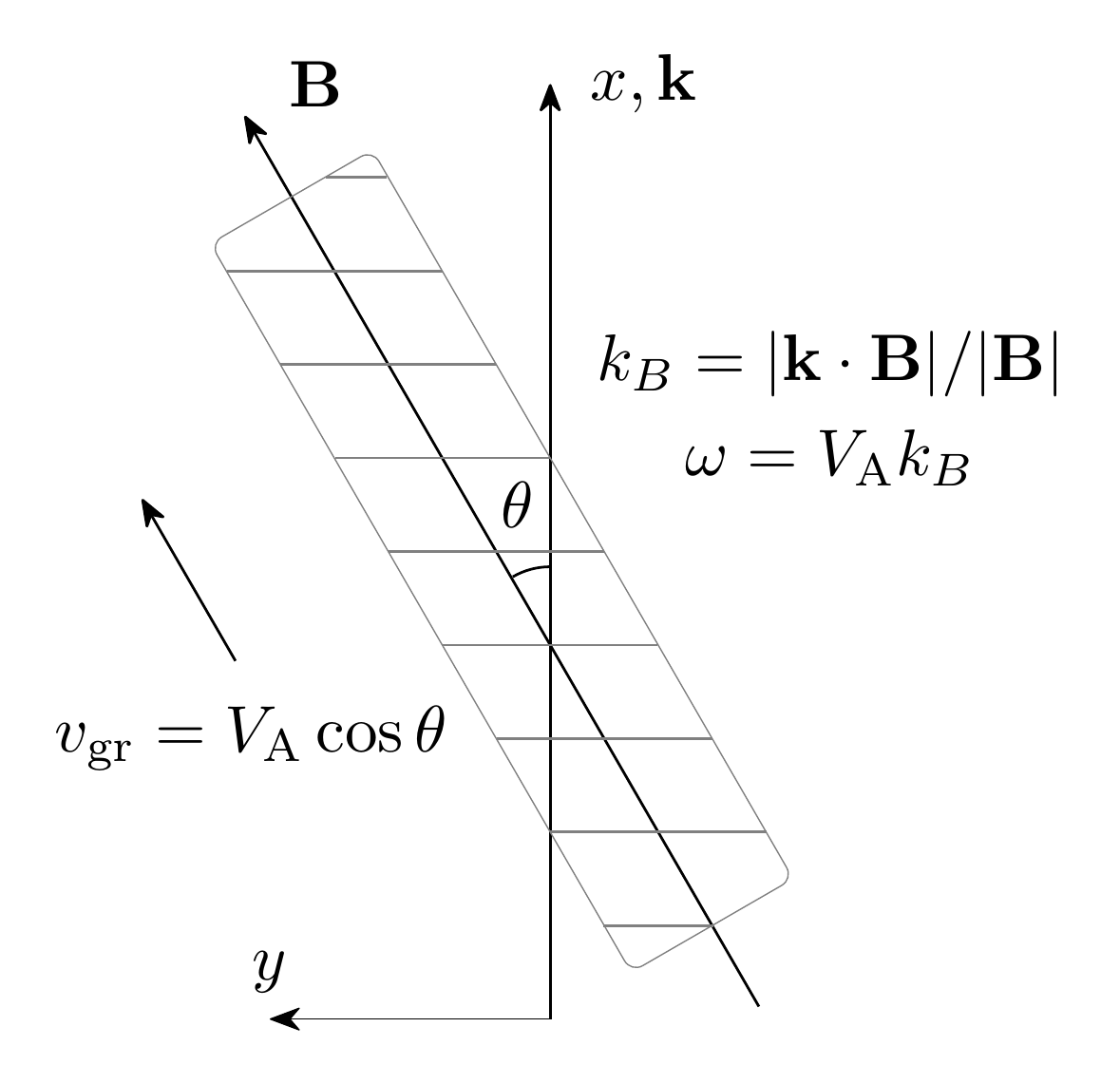}
\caption{Geometry of propagating of SMW in a constant external magnetic field $\mathbf{B}$.
The group velocity of the propagating wave packet $\mathbf{v}_\mathrm{gr}$ is along the external magnetic field.
The wave vector $\mathbf{k}$ is along the normal of wave fronts (equiphase planes) shown here with lines parallel to the $y$ axis.}
\label{Fig:SMW}
\end{figure}

Small amplitude MHD waves
we treat as small variations 
of the magnetic field $\mathbf{b}$,
density $\rho^\prime$,
pressure $p^\prime$ and temperature $T^\prime$
\beal
& \mathbf{B}=\mathbf{B}_0+\mathbf{b},\quad
\rho=\rho_0+\rho^\prime,\\
& p=p_0+p^\prime,\quad
T=T_0+T^\prime
\end{align}
from their constant values.
The index $0$ we omit where it is obvious.
The variations of the pressure are related with variations 
of the density
\be
p^\prime\approx c_s^2\rho^\prime,\quad
\label{p_prime}
\ee
according the definition of the sound speed.
Here we recall also the equations of state
$pV^\gamma=\mathrm{const}$
and
$TV^{1/c_v}=\mathrm{const}$
for $S=\mathrm{const}$ (constant entropy $S$) and  
give the relations between
variations of the temperature, pressure and density\cite{LL5}
\be
\frac{T^\prime}{T}=\frac1{c_v}\frac{\rho^\prime}{\rho}
=\frac1{c_p}\frac{p^\prime}{p},\quad
\frac{\rho^\prime}{\rho}
=\frac{1}{\gamma}\frac{p^\prime}{p},\quad
\gamma\equiv\frac{c_p}{c_v}=\frac{5}{3}.
\ee
For a weak amplitude plane wave the 
variations of all variables are proportional to the imaginary exponent 
$\propto\e^{\mathrm{i}(\mathbf{k}\cdot\mathbf{r}-\omega t) }$
and phase velocity 
$u\equiv\omega/k$
the ratio between the frequency
$\omega$ and the modulus $k=|\mathbf{k}|$
of the wave-vector.
For a plane wave the time $t$ and space $\mathbf{r}$
derivatives are reduced to multiplication 
\be
\partial_t=-\mathrm{i}\omega,\qquad \nabla=\mathrm{i}\mathbf{k}
\ee
and the MHD equations,
omitting index $0$ and 
imaginary unit~$\mathrm{i}$,
reads\cite{LL8}
\beal
&
\label{Eq:vt}
-\omega \rho \mathbf{v}
=-c_s^2\mathbf{k}\rho^\prime
+\mathbf{B}\times
\left(\mathbf{k}\times\mathbf{b}\right)/\mu_0,
\quad \mu_0=4\pi,\\
&
-\omega \mathbf{b}=\mathbf{k}\times
\left(\mathbf{v}\times\mathbf{B}\right),
\quad
\omega\rho^\prime=\rho\mathbf{k}\cdot\mathbf{v},
\;\;\varepsilon_0=1/4\pi,
\label{rho_prime}
\end{align}
where $\mathbf{k}\cdot\mathbf{b}=0$.
We use Gaussian units but
in SI\&C:
$\mu_0\,*=10^{-7}$ and 
$\varepsilon_0 /=c^2\, 10^{-7}$,
i.e. all formulae in the present work are 
written in system invariant form.
In Heaviside–Lorentz units $\mu_0=1$ and $\varepsilon_0=1$.
The $x$-axis is chosen along the wave-vector
$\mathbf{k}=k\mathbf{e}_x$,
and $y$-axis is in the plane of the wave-vector
and the constant magnetic field
\begin{align}
\mathbf{B}=B_x \mathbf{e}_x
+B_y \mathbf{e}_y,\quad 
B_x=B\cos\theta,\quad B_y=B\sin\theta,
\nn
\end{align}
see Fig.~\ref{Fig:SMW}.
The unit vector along the external magnetic field is
\be
\mathbf{e}_B=\mathbf{B}/B
=\cos\theta\,\mathbf{e}_x
+\sin\theta\,\mathbf{e}_y
\ee
Dividing by $k$, 
the nonzero components of the 
MHD equations Eqs.~\ref{Eq:vt} and \ref{rho_prime} read
\beal
&
\rho u \left(1-\frac{c_s^2}{u^2}\right)\!v_x
=B_yb_y/\mu_0,\quad
\label{v_x}\\
&
\rho u v_y=-B_xb_y/\mu_0,
\label{v_y}\\
&
u b_y= B_y v_x-B_x v_y.
\end{align}
Expressing velocity components $v_x$ and $v_y$ 
from the first two equations and substituting it in the third one gives a quadratic equation for the phase velocity
$u=\omega/k$ which have the solutions
describing fast (f) and slow (s) magnetosonic waves\cite{LL8}
\be
u^2_\mathrm{f,s}=\frac12
\left\{V_\mathrm{A}^2+c_s^2\pm
\left[(V_\mathrm{A}^2+c_s^2)^2
-4c_\theta^2V_\mathrm{A}^2c_s^2\right]^{1/2}
\right\}.
\ee
For small magnetic fields for SMW wave we have
\be
u=u_s\approx V_\mathrm{A}c_\theta\ll c_s,
\qquad c_\theta\equiv|\cos\theta|,
\ee
where
\be
\mathbf{V}_\mathrm{A}\equiv \frac{\mathbf{B}}{\sqrt{\mu_0\rho}}, \quad
\rho V_\mathrm{A}^2=B^2/\mu_0
\ee
is the speed of AW and $u_\mathrm{A}\equiv V_\mathrm{A}c_\theta$
is the modulus of its projection along the $x$-axis.
In such a way for the dispersion of SMW we have
\beal
\omega=\left\vert \mathbf{V}_\mathrm{A}
\cdot\mathbf{k}\right|,\quad
\mathbf{v}_\mathrm{gr}
\equiv\frac{\partial \omega}{\partial\mathbf{k}}
=\mathbf{V}_\mathrm{A}\,\mathrm{sgn}
(\mathbf{B}\cdot\mathbf{k}).
\end{align}
The frequency of SMW can be expressed 
by the projection of Alfv\`en speed along the 
wave vector $u_\mathrm{A}$
\beal &
\omega=u_\mathrm{A}k
=V_\mathrm{A}k_B,\\
&
u\approx u_\mathrm{A}\equiv V_\mathrm{A}c_\theta
\ll c_s,
\quad k_B=kc_\theta
\end{align}
or by projection of the wave-vector along the magnetic field $k_B$.
For SMW the last inequality substituted in
Eqs.~(\ref{v_x}) and (\ref{v_y}) gives
\be
\left(\frac{c_s^2}{u^2}-1\right)\approx c_s^2/u^2 \gg 1,
\ee
and we have approximate expressions for the components
of the velocity
\beal&
v_y=-\frac{B_x}{\mu_0\rho u} b_y,\quad
\mathbf{k}\cdot\mathbf{v}
=-\frac{B_x k}{\mu_0\rho u} b_y
\label{v_y approx}
\\&
v_x\approx-\frac{u}{c_s^2}\frac{B_y}{\mu_0\rho}b_y,
\quad b_x=0.
\end{align}
Then from Eq.~(\ref{rho_prime}) we obtain the variation 
of the density
\be
\frac{\rho^\prime}{\rho}
=\frac{\mathbf{k}\cdot\mathbf{v}}{\omega}
=\frac{B_yb_y}{\mu_0 \rho c_s^2},
\ee
and from Eq.~(\ref{p_prime})
\be
p^\prime=c_s^2\rho^\prime=-\frac{B_yb_y}{\mu_0}
=-\frac{B}{\mu_0} s_\theta b_y,\quad 
s_\theta\equiv \sin(\theta),
\ee
we express the variations of the pressure
proportional to the small wave component of the magnetic field
\be
b_y=b_0\cos (kx-\omega t), \quad
\mathbf{b}=\mathbf{b}_0\cos (kx-\omega t).
\ee
The unit vector $\mathbf{e}_b\equiv \mathbf{b}_0/b_0=\mathbf{e}_y$
along the oscillating component of the magnetic field 
has angle $\pi/2-\theta$ with the constant one
\be
\left(\mathbf{e}_b\cdot \mathbf{e}_B\right)^2
=s_\theta^2.
\ee

Now we can express the averaged density of the wave
energy which according to the virial theorem is twice
the averaged density of the magnetic energy
\beal
\mathcal{E}=2\left<\frac{b^2}{2\mu_0} \right>
=\frac{b_0^2}{2\mu_0},\qquad 
\overline{\mathbf{q}}=\mathcal{E}\mathbf{v}_\mathrm{gr}
\label{q}
\end{align}
and the averaged density of the pressure oscillations
which is one important ingredient
of the forthcoming analysis 
\be
\left<(p^\prime)^2\right>
=\frac{B^2}{\mu_0}s_\theta^2\frac{b_0^2}{2\mu_0}
=\frac{B^2}{\mu_0}\,\mathcal{E}s_\theta^2.
\label{link}
\ee
We mention that the pressure oscillations disappear    
for wave-vector parallel to the magnetic field
($\mathbf{k} \parallel \mathbf{B}$ or $\sin\theta=0$) and
$\mathbf{v} \cdot \mathbf{k} = \mathbf{b} \cdot \mathbf{k} = 0$,
the waves are purely transverse.

After the consideration of dissipationless 
wave propagation in the next subsection
we recall the results for SMW damping.

%%%%%%%%%%%%%%
\subsection{MHD Absorption}

In the WKB approximation we suppose that 
wave amplitudes have small exponential decay
$\e^{-\gamma_t t}$
as a function of time or 
space extinction 
$\e^{-\gamma x}$
if we trace a traveling wave packet.

For the energy flux and density 
we have quadratic dependence $\e^{-2\gamma_x x}$
and the extinction 
\be
\gamma_x=
\frac{\mathcal{\overline{Q}_\mathrm{MHD}}}
{2\overline{q}_x}
\ee
is given by the ratio of
the time averaged power of MHD dissipation
$\overline{Q}_\mathrm{MHD}$
and the energy flux $\overline{q}_x$.\cite{LL8}
In dissipationless approximation the substitution of
\be
b_y=b_0\cos(kx-\omega t),
\quad b_x=b_z=0
\ee
and the derived velocity $v_y$ Eq.~(\ref{v_y approx})
in the formula for the Pointing vector in MHD
\be
\mathbf{q}\approx\mathbf{S}
\approx\mathbf{B}\times
\left(\mathbf{v}\times\mathbf{B}\right)\!/\mu_0
\ee
gives
\beal&
\overline{\mathbf{S}}
=\mathbf{V}_\mathrm{A}\mathcal{E},\\&
\mathcal{E}=
\left<\mathbf{b}^2/2\mu_0+\rho\mathbf{v}^2/2\right>
=b_0^2/2\mu_0,
\end{align}
in agreement with Eq.~(\ref{q}).
For the $x$-component we have
\be
\overline{q}_x=u_\mathrm{A}\mathcal{E}.
\ee
The small dissipation is proportional to the dissipative coefficients
\beal
\mathcal{Q}_\mathrm{MHD}
=\nu_\mathrm{k}\rho
\left(\frac{\partial \mathbf{v}}{\partial x}\right)^{\!\!2}
+\nu_\mathrm{m}
\left(\frac{\partial \mathbf{b}}{\partial x}\right)^{\!\!2}
\!\!/\mu_0
\end{align}
paramererized by kinematic $\nu_\mathrm{k}=\eta/\rho$
and magnetic diffusivity $\nu_\mathrm{m}=\varepsilon_0c^2\varrho$,
where $\eta$ is viscosity coefficient 
and  $\varrho$ is the Ohmic resistivity.
Expressing $v_y$ from Eq.~(\ref{v_y})
and assuming $v_x\approx 0$ from Eq.~(\ref{v_x}) meaning that
$\mathrm{div}\,\mathbf{v}\approx0$
after some algebra we obtain
\beal&
\gamma_x=\frac{\gamma_t}{u_a},
\quad \gamma_t
=\frac12 (\nu_\mathrm{k}+\nu_\mathrm{m})k^2.
\end{align}

If we trace a wave packet of SMW propagating along
magnetic force lines $\mathbf{B}$
at distance $l=x/c_\theta$ 
for the energy damping $\propto\e^{-2\gamma_l l}$
we have the extinction
\be
\gamma_l
=
\frac{\mathcal{\overline{Q}_\mathrm{MHD}}}
{2V_\mathrm{A}\mathcal{E}}
=\frac{\gamma_t}{V_\mathrm{A}}
=\frac{(\nu_\mathrm{k}+\nu_\mathrm{m})k^2}
{2V_\mathrm{A}}.
\label{gamma_MHD}
\ee
This space damping rate does not depend on the angle 
$\theta$.
For AW we have velocity $v_z$ and magnetic field 
$b_z$ oscillations only normal to the 
($\mathbf{k}$-$\mathbf{B}$) plane direction but for small 
magnetic field $V_\mathrm{A}\ll c_s$ 
the dispersion and wave damping are the same.
The difference appears  when we analyze the 
chemical damping of SMW.

After this recall of the well-known result we 
analyze in the next section the chemical damping.

%%%%%%%%%%%%%%%%%%%%%%
\section{Ionization-recombination absorption}
   \label{Ionization-recombination absorption}

The degree of the ionization
\be
\alpha=\frac{n_p}{n_\rho}
\ee
is a result of the continuous balance
of ionization and recombination processes
\be
\frac{\mathrm{d}n_p}{\mathrm{d}t}
=\beta n_0n_e-\gamma_\mathrm{rec} n_pn_e^2
\ee
with temperature dependent rates of 
electron impact ionization 
$\beta(T)$
and two electron recombination 
$\gamma(T).$
For dense enough plasma the radiative processes 
have negligible contribution, especially for optically thin
plasma regions.\cite{Pitaevskii:62,LL10}

In thermal equilibrium the degree of ionization
is given by Saha equation\cite{Saha,LL5}
\be
\frac{\bar n_p \bar n_e}{\bar n_0 }
=n_{_\mathrm{S}} 
\equiv \left ( \frac{mT}{2 \pi \hbar^2} \right )^{3/2}
\e^{-I/T},
\label{balance}
\ee
where $I$ is the ionization energy.
The rates of ionization and recombination processes 
in equilibrium are equal
\be
\nu=\beta \bar n_e \bar n_0
=\gamma_\mathrm{rec} \bar n_e^2 \bar n_p.
\ee
The variable $\nu(T)$ gives the number of reactions 
\begin{align}
\beta: \quad & \mathrm{H} + \mathrm{e} \longrightarrow \mathrm{p} + \mathrm{e} + \mathrm{e}, \\
\gamma_\mathrm{rec} : \quad & \mathrm{p} +\mathrm{e} + 
\mathrm{e} \longrightarrow \mathrm{H} + \mathrm{e}
\end{align}
per unit volume and unit time.
From this rate one can create a temperature dependent
variable 
\be
Q_{\iota}\equiv T\nu
\ee
with dimension of power density; energy per unit volume and unit time.

When MHD waves propagate through the plasma 
oscillations of the pressure $p^\prime$, 
density $\rho^\prime$ and the temperature $T^\prime$
perturbate the chemical equilibrium and induce
variations of the chemical composition and ionization degree $\alpha$.
This extra chemical chaos creates an additional mechanism of increasing of entropy and wave energy dissipation
\be
\mathcal{Q}_\mathrm{ion}=Q_{\iota} \!
\left<\chi^2\right>,
\ee
where brackets denotes wave period averaging.

The main detail of the chemical energy dissipation
is the deviation from the chemical equilibrium 
\begin{align}
\chi \equiv\frac{n_e n_p}{n_0n_{_\mathbf{S}}}-1
=\frac{n_e n_p}{n_0}
\frac{\bar n_0}{\bar n_e \bar n_p}-1
\label{chi_nuovo}
\end{align}
described in detail in a recent 
Ref.~\onlinecite{var_mish_2020}.
We suppose that the variations of the chemical
chomposition are relatively small, 
and the frequency of SMW is high enough
\begin{align}&
\omega\gg \alpha (1-\alpha) \beta n_\rho,\\&
\bar n_e= \bar n_p
=\alpha \bar n_\rho,\quad \bar n_0=(1-\alpha)n_\rho.
\end{align}
In equilibrium $\bar\chi=0$ and we have to calculate
the small small change of
the variable $\chi$ describing 
the deviation from the chemical equilibrium
substituting in Eq.~(\ref{chi_nuovo})
all necessary details
\beal&
n_e=\bar n_e+n_e^\prime,\quad
n_p=\bar n_p+n_p^\prime,\quad
n_0=\bar n_0+n_0^\prime,\quad\\&
n_{_\mathbf{S}}(T+T^\prime)=
n_{_\mathbf{S}}+n_{_\mathbf{S}}^\prime
=n_{_\mathbf{S}}(T)
+\frac{\mathrm{d}
n_{_\mathbf{S}}}{\mathrm{d}T}\,T^\prime.
\end{align}
For linearized waves and small $|\chi|\ll1$
we have
\beal
\chi\approx
\frac{n_e^\prime}{n_e}
+\frac{n_p^\prime}{n_p}
-\frac{n_0^\prime}{n_0}
-\frac{n_{_\mathbf{S}}^\prime}{n_{_\mathbf{S}}}.
\end{align}
All relative changes of the variables can be expressed
by the relative change of the pressure
\beal
\frac{n_e^\prime}{n_e}
=\frac{n_p^\prime}{n_p}
=\frac{n_0^\prime}{n_0}
=\frac{\rho^\prime}{\rho}
=\frac{1}{\gamma}\frac{p^\prime}{p}
\end{align}
and the Saha density
\beal
\frac{n_{_\mathbf{S}}^\prime}{n_{_\mathbf{S}}}
=\frac{\mathrm{d}n_{_\mathbf{S}}}
{n_{_\mathbf{S}}\mathrm{d}T}T^\prime
=\left(\frac{I}{T}+c_v\right)
\frac{T^\prime}{T}
=\left(\frac{I}{T}+c_v\right)
\frac{p^\prime}{c_pp}.
\end{align}
Due to detailed text-book recalling of the 
SMW dynamics we easily arrive at a simple 
result
\beal 
\chi=\left(\frac{I}{c_pT}+\frac2{\gamma}\right)
\,\frac{p^\prime}{p}
\end{align}
and its  square can be easily averaged using Eq.~(\ref{link})
\beal 
\left<\chi^2\right>
&=\left(\frac{I}{c_pT}+\frac2{\gamma}\right)^{\!\!2}
\,\frac{\left<(p^\prime)^2\right>}{p^2}\\
&=\left(\frac{I}{c_pT}+\frac2{\gamma}\right)^{\!\!2}
\frac{B^2}{\mu_0p}\frac{\mathcal{E}}{p}\,s_\theta^2.
\end{align}
Multiplying with the power density rate 
we finally derive
the main result of the present work:
the mean energy dissipation of a SMW
propagating in magnetized plasma
\beal
\mathcal{Q}_\mathrm{ion}
=Q_{\iota}
\left(\frac{I}{c_pT}+\frac2{\gamma}\right)^{\!\!2}
\frac{B^2}{\mu_0p}\frac{\mathcal{E}}{p}\,s_\theta^2.
\end{align}
Now for the time damping we obtain
\beal&
\tilde\gamma_t
=\frac{\mathcal{Q}_\mathrm{ion}}{2\mathcal{E}}
=\frac{Q_{\iota}}{p}
\left(\frac{I}{c_pT}+\frac2{\gamma}\right)^{\!\!2}
\frac{B^2/2\mu_0}{p}\,s_\theta^2,\\&
\end{align}
and for the extinction at low temperatures $T\ll I$
we have an additional chemical term
\beal&
\tilde\gamma_l
\approx\frac{\mathcal{Q}_\mathrm{ion}}
{2\mathcal{E}V_\mathrm{A}}
=\frac{Q_{\iota}}{pV_\mathrm{A}}
\left(\frac{I}{c_pT}\right)^{\!\!2}
\frac{B^2/2\mu_0}{p}\,s_\theta^2,
\label{gamma_chem}
\end{align}
which disappears at small angles $\theta\ll 1.$
In the next final section we will discuss
the difference between two damping mechanisms
giving total SMW extinction
\be
\gamma_\mathrm{tot}=\gamma_l+\tilde\gamma_l.
\ee

%%%%%%%%%%%%%%%%%%
\section{Discussion and conclusions}
   \label{Discussion and conclusions}

The angular dependence of the chemical damping
obtained in Eq.~(\ref{gamma_chem})
$\tilde\gamma_l\propto \sin^2\theta$ 
is the main difference between the chemical damping and
the MHD one.
Here we wish to emphasize also that the derived
new ionization-recombination damping 
is frequency independent and has no hydrodynamic 
sense as second viscosity, for example.
The MHD damping according to
Eq.~(\ref{gamma_MHD})
$\gamma_{l}\propto k^2\propto \omega^2$
is proportional to the square of the wave-vector 
and square frequency.

Roughly speaking MHD damping is a low pass filter
while ionization-recombination mechanism 
is a bass damper.

Imagine that turbulence generates 
broad distribution of MHD waves and the 
angular distribution of the spectral density is almost
constant at small angles between the wave-vector
and constant magnetic field
\be
\cos\theta=\frac{\mathbf{k}\cdot\mathbf{B}}{kB}.
\ee
If then SMW pass through a partially ionized region
with length $l$ the chemical damping gives
transmission coefficient 
\beal&
\tilde T_\mathrm{SMW}=\e^{-2\tilde\gamma_l l}
=\exp\left(-\frac{\theta^2}{2\theta_0^2}\right),
\\&
\frac{1}{2\theta_0^2}
=\frac{2\nu Tl}{pV_\mathrm{A}}
\left(\frac{I}{c_pT}\right)^{\!\!2}
\frac{B^2/2\mu_0}{p}\,s_\theta^2,\\&
\theta_0=\frac{c_pT}{2I}
\sqrt{\frac{pV_\mathrm{A}\beta_\mathrm{pl}}
{\nu Tl}}\ll1,\quad
\beta_\mathrm{pl}\equiv\frac{p}{p_B},\\&
p_B=B^2/2\mu_0,\quad 
p=(\bar n_e+\bar n_p+\bar n_0)T.
\end{align}
In other words, 
strong ionization-recombination absorption 
gives a cumulative small angle distribution
of SMW. 
Waves with significant angles $\theta$ are absorbed
and dominantly AW will pass through large area of partially 
ionized plasma.

How this can be checked by observations.
Imagine that in an observation point 
we have a good record of the time dependence of
the magnetic field $\mathbf{B}(t)$.
Time averaging can give mean value $B_0$
and orientation $\mathbf{e}_B$ of the constant component of the magnetic field
\be
\mathbf{B}_0=\left<\mathbf{B}(t)\right>,\quad
\mathbf{e}_B=
\mathbf{B}_0/|\mathbf{B}_0|.
\nn
\ee
Then we can make Fourier analysis and calculate the
wave components of the magnetic field
for all frequencies $\omega$
\begin{align}&
\mathbf{b}_\omega^\prime=
\left<\left(\mathbf{B}(t)-\mathbf{B}_0\right)
\cos(\omega t)\right>,\quad
\mathbf{e}_\omega^\prime
=\mathbf{b}_\omega^\prime
/|\mathbf{b}_\omega^\prime|, \nn \\&
\mathbf{b}_\omega^{\prime\prime}=
\left<\left(\mathbf{B}(t)-\mathbf{B}_0\right)
\sin(\omega t)\right>,\quad
\mathbf{e}_\omega^{\prime\prime}
=\mathbf{b}_\omega^{\prime\prime}
/|\mathbf{b}_\omega^{\prime\prime}|. \nn
\end{align}
For the considered in Sec.~\ref{Recalling SMW} example
we have
\be
\mathbf{e}_B
=\cos\theta\mathbf{e_x}
+\sin\theta\mathbf{e_y},\quad
\mathbf{e}_\omega^\prime=\mathbf{e_y},
\quad 
\mathbf{e}_B\cdot\mathbf{e}_\omega^\prime
=\sin\theta.
\nn
\ee
In the general case we have
different angles for all Fourier frequencies
\be
\sin(\theta_\omega^\prime)=
\mathbf{e}_B\cdot\mathbf{e}_\omega^\prime,
\quad
\sin(\theta_\omega^{\prime\prime})=
\mathbf{e}_B\cdot
\mathbf{e}_\omega^{\prime\prime},
\nn
\ee
and it is worthwhile the study
probability distribution function of angles $\theta$.
Our simple consideration predicts
Gaussian distribution 
\be
P(\theta)\propto \exp(-\theta^2/2\theta_0^2)
\ee
created by long regions of partially ionized
plasma.
Roughly speaking SMW filtered by large regions of partially ionized plasmas 
will be almost transverse.

Every similarity with phenomena in the magnetized 
atmosphere even in the nearest star is random.
We present a purely academic study.

Last but not least the ionization rate $\beta=\left<v\sigma\right>$
is given by the Maxwell velocity $v$ averaging of the 
electron impact ionization cross-section $\sigma$ 
and all details of the proposed new damping mechanisms 
are well studied.

The considered in the present work the chemical damping of the pressure 
oscillations in some sense belongs to the notions
of plasma multi-fluid approach.
Not only relative velocity between different 
fluids\cite{Zaqarashvili:11,Ballester:18}
creates dissipation as a friction forces.
Periodic oscillations around the Saha equilibium 
for some MHD modes of partially ionized plasmas 
can give even bigger dissipation and indispensably have to be taken into account in the arsenal of the plasma physics notions.

%%%%%%%%%%%%%%%%
\section*{Acknowledgments} 
The authors appreciate stimulating discussions correspondence with 
Dantchi Koulova, 
Kamen~Kozarev, Hassan~Chamati, Yavor~Boradjiev, Nedko~Ivanov, and Stanislav~Varbev.
%  This work is partially supported by \dots

\section*{Data Availability Statement}

Data sharing is not applicable to this article
as no new data were created or analyzed in this study,
which is a purely theoretical one.

%%%%%%%%%%%%%%%%


\begin{thebibliography}{99}

\bibitem{Alfven:42}
H.~Alfv\'en,
``Existence of Electromagnetic-Hydrodynamic Waves'',
Nat. \textbf{150}, 405 (1942).
	
\bibitem{Alfven:47}
H.~Alfv\'en, 
``Granulation, magnetohydrodynamic waves, and the heating of the solar corona'', 
Mon. Not. Roy. Astr. Soc. \textbf{107}, 211 (1947).

\bibitem{LL8}         
L.~D.~Landau and E.~M.~Lifshitz,	
\textit{Electrodynamics in Continuous Media} in 
L.~D.~Landau and E.~M.~Lifshitz,	
\textit{Course of Theoretical Physics, Vol.~VIII}
(Pergamon Press, New York, 1960),
Sec.~52 ``Hydromagnetic waves''.

\bibitem{Mond:94}
I. M. Rutkevich and M. Mond,
``Localization of slow magnetosonic waves in the solar corona'',
Phys. Plasmas \textbf{3}(1), 392 (1996).%; doi: 10.1063/1.871865

\bibitem{Campos:99}
L. M. B. C. Campos,
``On the viscous and resistive dissipation of magnetohydrodynamic waves'',
Phys. Plasmas \textbf{6}(1), 57 (1999).

\bibitem{Wijn:07}
A.~G.~de~Wijn, B.~De~Pontieu and R.~J.~Rutten,
``Chromospheric and Transition-Region Dynamics in Plage'' in 
\textit{The Physics of Chromospheric Plasmas, 9-13 October, 2006, Coimbra, Portugal},
ed. by P.~Heinzel, I.~Dorotovi\u{c}, and R.~J.~Rutten,
(Astronomical Society of the Pacific Conference Series \textbf{368}, San Francisco, 2007);\\
``Fourier Analysis of Active-region Plage'',
Astrophys. J \textbf{654}, 1128 (2007).

\bibitem{Ruderman:12}
R.~J.~Morton, G.~Verth, D.~B.~Jess, D.~Kuridze, M.~S.~Ruderman, M.~Mathioudakis and R.~Erd\'elyi,
``Observations of ubiquitous compressive waves in the Sun’s chromosphere'',
Nat Commun \textbf{3}, 1315 (2012).
%DOI: 10.1038/ncomms2324

\bibitem{Maneva:17}
Y.~G.~Maneva, A.~A.~Laguna, A.~Lani and S.~Poedts,
``Multi-fluid Modeling of Magnetosonic Wave Propagation in the Solar Chromosphere:
Effects of Impact Ionization and Radiative Recombination'',
Astrophys. J \textbf{836}, 197 (2017).

\bibitem{Zhelyazkov:87}
W.~Sahyouni  Zh.~Kiss’ovski and I.~Zhelyazkov,
``Chromospheric and Coronal Heating Due to the Radiation and Collisional Damping of Fast Magnetosonic Surface Waves'',
Z. Naturforsch. A \textbf{42a}(12), 1443 (1987).

\bibitem{Hristov:11}
Z.~D.~Dimitrov, Y.~G.~Maneva, T.~S.~Hristov and T.~M.~Mishonov, 
``Over-reflection of slow magnetosonic waves by homogeneous shear flow: Analytical solution'',
Phys. Plasmas \textbf{18}, 082110 (2011).
%https://doi.org/10.1063/1.3624776

\bibitem{Gogoberidze:04}
G.~Gogoberidze, G.~D.~Chagelishvili, R.~Z.~Sagdeev, and D.~G.~Lominadze,
``Linear coupling and overreflection phenomena of magnetohydrodynamic
waves in smooth shear flows'',
Phys. Plasmas \textbf{11}, 4672 (2004).%; doi: 10.1063/1.1789998

\bibitem{Ballester:18}
J.~L.~Ballester, I.~Alexeev, M.~Collados, T.~Downes, R.~F.~Pfaff, H.~Gilbert, M.~Khodachenko, E.~Khomenko, I.~F.~Shaikhislamov, R.~Soler, E.~V\'azquez-Semadeni, T.~Zaqarashvili,
``Partially Ionized Plasmas in Astrophysics'',
Space Sci. Rev. \textbf{214}, 58 (2018).
%https://doi.org/10.1007/s11214-018-0485-6

\bibitem{Perelomova:20}
Anna Perelomova,
``On description of periodic magnetosonic perturbations in a quasi-isentropic plasma with
mechanical and thermal losses and electrical resistivity''
Phys. Plasmas \textbf{27}, 032110 (2020).%; doi: 10.1063/1.5142608

\bibitem{Zaqarashvili:11}
T.~V.~Zaqarashvili, M.~L.~Khodachenko, and H.~O.~Rucker,
``Magnetohydrodynamic waves in solar partially ionized plasmas:
two-fluid approach'',
Astron. Astrophys. \textbf{529}, A82 (2011).% DOI: 10.1051/0004-6361/201016326

\bibitem{LL5}
L.~D.~Landau, E.~M.~Lifshitz and L.~P.~Pitaevskii,
\textit{Statistical Physics Part 1} in
L.~D.~Landau and E.~M.~Lifshitz,
\textit{Landau-Lifshitz Course on Theoretical Physics, Vol.~V}
(3rd ed., Pergamon Press, New York, 1980), 
Sec.~43, ``Ideal gas with constant heat capacity'',
Sec.~45, ``Mono-atomic gas'',
Sec.~46, ``Mono-atomic gas. Influence of electronic momentum'', Eq.~(46.1a),
Sec.~104, ``Ionization equilibrium''.

\bibitem{Pitaevskii:62}
L.~P.~Pitaevskii, ``Electron Recombination in a Monatomic gas'', JETP \textbf{15}(5), 919 (1962).

\bibitem{LL10}
E.~M.~Lifshitz and L.~P.~Pitaevskii,
\textit{Physical Kinetics} in
L.~D.~Landau and E.~M.~Lifshitz,
\textit{Landau-Lifshitz Course on Theoretical Physics, Vol.~X}
(Pergamon, New York, 2002), Sec.~24, ``Recombination and ionization''.

\bibitem{Saha}
%%M.~N.~Saha, ``LIII. Ionization in the solar chromosphere'',
%%Phil. Mag. Ser. 6, \textbf{40}(238), 472-488 (1920);
%%
M.~N.~Saha, ``On a physical theory of stellar spectra'', 
Proc. R. Soc. Lond. A, \textbf{99}(697), 135-153 (1921).

\bibitem{var_mish_2020}
T.~M.~Mishonov, A.~M.~Varonov,
``Sound Absorption in Partially Ionized Hydrogen Plasma and Heating Mechanism of Solar Chromosphere'',
arXiv:2005.05056 [physics.plasm-ph].

\end{thebibliography}
\end{document}